\RequirePackage{fix-cm} 
\documentclass[a4paper, twoside, reqno, dvips, 12pt]{amsart}
\usepackage{fixltx2e}   

\usepackage[latin1]{inputenc}
\usepackage[T1]{fontenc}

\usepackage{eucal}
\usepackage{esint}
\usepackage{dsfont}
\usepackage{xspace}
\usepackage{amsgen}
\usepackage{amsthm}
\usepackage{amssymb}
\usepackage{amsmath}
\usepackage{upgreek}
\usepackage{amsfonts}
\usepackage{MnSymbol}
\usepackage{mathrsfs}
\usepackage{stmaryrd}
\usepackage{mathtools}
\usepackage[nice]{nicefrac}

\usepackage{a4wide}

\usepackage[dvipsnames, table]{xcolor}

\definecolor{oneblue}{rgb}{0.0, 0.0, 0.85}
\definecolor{bluepigment}{rgb}{0.2, 0.2, 0.6}
\definecolor{darkgrey}{rgb}{0.273, 0.281, 0.30}
\definecolor{Lightgray}{rgb}{0.89, 0.89, 0.89}
\definecolor{Lightblue}{RGB}{214, 214, 214}
\definecolor{bckg}{RGB}{20.8, 20.8, 20.8} 
\definecolor{charcoal}{rgb}{0.21, 0.27, 0.31}
\definecolor{darkelectricblue}{rgb}{0.33, 0.41, 0.47}

\usepackage[sort&compress, comma, square, numbers]{natbib}

\usepackage{psfrag}
\usepackage{graphicx}
\usepackage{subfigure}
\usepackage{morefloats}
\usepackage{indentfirst}

\usepackage{acronym}
\usepackage{microtype}
\usepackage[labelsep=period,%
            labelfont={bf,sf,color=bluepigment},%
            justification=raggedright]{caption}

\usepackage[perpage, symbol]{footmisc}

\usepackage[usenames, dvipsnames, pdf]{pstricks}
\usepackage{epsfig}
\usepackage{pst-grad} 
\usepackage{pst-plot} 

\usepackage[colorlinks,
           urlcolor=oneblue,
           linkcolor=oneblue,
           citecolor=NavyBlue,
           bookmarksopen=false,
           pdfpagemode=UseNone,
           pagebackref]{hyperref}

\usepackage[explicit]{titlesec}

\titleformat{\section}
  {\color{NavyBlue}\Large\sffamily\bfseries}
  {}
  {0em}
  {\colorbox{bckg!5}{\parbox{\dimexpr\linewidth-2\fboxsep\relax}{\centering\thesection. #1}}}
  [\vspace*{0.33em}]

\titleformat{name=\section,numberless}
  {\color{NavyBlue}\Large\sffamily\bfseries}
  {}
  {0.0em}
  {\colorbox{bckg!10}{\parbox{\dimexpr\linewidth-2\fboxsep\relax}{\centering#1}}}
  [\vspace*{0.33em}]

\titleformat{\subsection}
  {\color{NavyBlue}\large\sffamily\bfseries}
  {}
  {0.0em}
  {\colorbox{bckg!5}{\parbox{\dimexpr\linewidth-2\fboxsep\relax}{\centering\thesubsection. #1}}}
  [\vspace*{0.33em}]

\titleformat{name=\subsection,numberless}
  {\color{NavyBlue}\Large\sffamily\bfseries}
  {}
  {0em}
  {\colorbox{bckg!10}{\parbox{\dimexpr\linewidth-2\fboxsep\relax}{\centering#1}}}
  [\vspace*{0.33em}]

\titleformat{\subsubsection}
  {\color{bluepigment}\sffamily\normalsize\bfseries}
  {\thesubsubsection}
  {0.5em}
  {#1}
  [\vspace*{0.33em}]

\titleformat{\paragraph}[runin]
  {\color{bluepigment}\sffamily\small\bfseries}
  {}
  {0em}
  {#1}

\titlespacing{\section}{1.0em}{1.5em plus 2pt minus 2pt}%
{1.0em plus 2pt minus 2pt}[0em]
\titlespacing{\subsection}{1.0em}{1.5em plus 2pt minus 2pt}%
{1.0em}[0em]
\titlespacing{\subsubsection}{1.0em}{1.5em plus 2pt minus 2pt}%
{1.0em plus 2pt minus 2pt}[0em]

\usepackage{titletoc}

\contentsmargin{0.5em}
\newlength{\tocsep} 
\titlecontents{section}[\tocsep]
  {\addvspace{10pt}\bfseries\sffamily}
  {\contentslabel[\thecontentslabel]{\tocsep}}
  {}
  {\ \titlerule*[0.75pc]{.}\ \thecontentspage}
  []
\titlecontents{subsection}[\tocsep]
  {\addvspace{8pt}\sffamily}
  {\contentslabel[\thecontentslabel]{\tocsep}}
  {}
  {\ \titlerule*[0.5pc]{.}\ \thecontentspage}
  []
\titlecontents*{subsubsection}[\tocsep]
  {\addvspace{2pt}\footnotesize\sffamily}
  {}
  {}
  {\ \titlerule*[0.35pc]{.}\ \thecontentspage}
  [\\*]

\makeatletter
\def\@setauthors{%
  \begingroup
  \def\thanks{\protect\thanks@warning}%
  \trivlist
  \centering\footnotesize \@topsep30\p@\relax
  \advance\@topsep by -\baselineskip
  \item\relax
  \author@andify\authors
  \def\\{\protect\linebreak}%
  \textsc{\normalsize\textcolor{darkelectricblue}{\authors}}%
  \ifx\@empty\contribs
  \else
    ,\penalty-3 \space \@setcontribs
    \@closetoccontribs
  \fi
  \endtrivlist
  \endgroup
}
\def\@settitle{\begin{center}%
  \baselineskip14\p@\relax
    \bfseries
    \textsc{\Large\textcolor{charcoal}{\@title}}
  \end{center}%
}
\makeatother

\usepackage{enumitem}
\setlist[description]{%
  topsep=30pt,               
  itemsep=5pt,               
  font={\bfseries\sffamily\color{NavyBlue}}, 
}

\usepackage{fancyhdr}
\usepackage{lastpage}

\newcommand*\Title{\textcolor{bluepigment}{Incoherent soliton ensembles}}
\newcommand*\Authors{\textcolor{bluepigment}{F.~Carbone, D.~Dutykh \& G.~El}}
\newcommand*{\plogo}{\textcolor{gray}{{\texttt{arXiv.org} / \textsc{hal}}}} 

\numberwithin{equation}{section}

\newcommand{\uh}{\hat{u}}

\newcommand{\N}{\mathds{N}}

\newcommand{\ud}{\mathrm{d}}
\newcommand{\ui}{\mathrm{i}}
\newcommand{\ue}{\mathrm{e}}
\newcommand{\F}{\mathcal{F}}

\newcommand{\Nn}{\mathscr{N}}

\renewcommand{\beta}{\upbeta}

\renewcommand{\leq}{\leqslant}
\renewcommand{\geq}{\geqslant}

\renewcommand{\L}{\mathcal{L}}

\newcommand{\const}{\mathrm{const}}

\newcommand{\cf}{\emph{cf.}\/}
\newcommand{\ie}{\emph{i.e.}\/}
\newcommand{\eg}{\emph{e.g.}\/}
\newcommand{\etc}{\emph{etc.}\/}

\newcommand{\pd}[2]{\frac{\partial #1}{\partial\/ #2}}

\usepackage{acronym}
\acrodef{bvp}[BVP]{Boundary Value Problem}
\acrodef{NSWE}{Nonlinear Shallow Water Equations}

\begin{document}

\title[\Title]{Macroscopic dynamics of incoherent soliton ensembles: soliton-gas kinetics and direct numerical modeling}

\author[F.~Carbone]{Francesco Carbone}
\address{CNR-IIA, Istituto Inquinamento Atmosferico, CNR, U.O.S. di Rende, c/o: UNICAL-Polifunzionale, 87036 Rende, Italy}
\email{f.carbone@iia.cnr.it}

\author[D.~Dutykh]{Denys Dutykh}
\address{LAMA, UMR 5127 CNRS, Universit\'e Savoie Mont Blanc, Campus Scientifique, 73376 Le Bourget-du-Lac Cedex, France}
\email{Denys.Dutykh@univ-savoie.fr}
\urladdr{http://www.denys-dutykh.com/}

\author[G.~El]{Gennady El$^*$}
\address{Department of Mathematical Sciences, Loughborough University, Loughborough, LE11 3TU, UK}
\email{G.El@lboro.ac.uk}
\thanks{$^*$ Corresponding author}



\begin{titlepage}
\thispagestyle{empty} 
\noindent
{\Large Francesco \textsc{Carbone}}\\
{\it\textcolor{gray}{CNR-IIA, U.O.S. di Rende, Italy}}
\\[0.02\textheight]
{\Large Denys \textsc{Dutykh}}\\
{\it\textcolor{gray}{CNRS--LAMA, Universit\'e Savoie Mont Blanc, France}}
\\[0.02\textheight]
{\Large Gennady \textsc{El}}\\
{\it\textcolor{gray}{Loughborough University, UK}}
\\[0.16\textheight]

\vspace*{0.65cm}

\colorbox{Lightblue}{
  \parbox[t]{1.0\textwidth}{
    \centering\huge\sc
    \vspace*{0.99cm}
    
    \textcolor{bluepigment}{Macroscopic dynamics of incoherent soliton ensembles: soliton-gas kinetics and direct numerical modeling}
    
    \vspace*{0.7cm}
  }
}

\vfill 

\raggedleft     
{\large \plogo} 
\end{titlepage}


\newpage
\maketitle
\thispagestyle{empty}


\begin{abstract}
We undertake a detailed comparison of the results of direct numerical simulations of the integrable soliton gas dynamics with the analytical predictions inferred from the exact solutions of the relevant kinetic equation for solitons. We use the KdV soliton gas as a simplest analytically accessible model yielding major insight into the general properties of soliton gases in integrable systems. Two model problems are considered: ({\it i}) the propagation of a `\emph{trial}' soliton through a one-component `\emph{cold}' soliton gas consisting of randomly distributed solitons of approximately the same amplitude; and ({\it ii}) collision of two cold soliton gases of different amplitudes (soliton gas shock tube problem) leading to the formation of an incoherend dispersive shock wave. In both cases excellent agreement is observed between the analytical predictions of the soliton gas kinetics and the direct  numerical simulations. Our results confirm relevance of the kinetic equation for solitons as a quantitatively accurate model for macroscopic non-equilibrium dynamics of incoherent soliton ensembles.

\bigskip
\noindent \textbf{\keywordsname:} soliton gas; Korteweg--de Vries equation; integrable turbulence; kinetic equation; direct numerical simulations


\end{abstract}


\newpage
\tableofcontents
\thispagestyle{empty}


\newpage
\section{Introduction}

Dynamics of incoherent nonlinear dispersive waves have been the subject of very active research in nonlinear physics for several decades, most notably in the contexts of ocean wave dynamics and nonlinear optics (see \eg \cite{Osborne2010, Laurie2012, Picozzi2014}). Two major areas where statistical properties of random ensembles of nonlinear waves play essential role are wave turbulence and rogue wave studies (see \cite{Kharif2009, Nazarenko2011} and references therein).

A very recent direction in the statistical theory of nonlinear dispersive waves introduced by V.E.~\textsc{Zakharov} (2009) \cite{Zakharov2009a} is \emph{turbulence in integrable systems}. It was suggested in \cite{Zakharov2009a} that many questions pertinent to a turbulent motion can be formulated in the framework of completely integrable systems. Such an `\emph{integrable turbulence}' theory has two natural premises: ({\it i}) nonlinear wave phenomena are often so complex that they must be described in statistical terms; ({\it ii}) integrable systems capture essential properties of many wave processes occurring in the real-world systems. Physical relevance of integrable turbulence theory was recently demonstrated in the fibre optics experiments \cite{Randoux2014, Walczak2015}.

Solitons play the key role in the characterisation of nonlinear wave fields in dispersive media, therefore the theory of \emph{soliton gases} in integrable comprises an important part of the general theory of integrable turbulence \cite{Zakharov2009a}. The very recent observations of dense statistical ensembles of solitons in shallow water wind waves in the ocean \cite{Costa2014} well modelled by the KdV equation and in the laminar/turbulent transition in fibre lasers \cite{Turitsyna2013} described by the defocusing Nonlinear Schr\"odinger (NLS) equation provide further physical motivation for the development of the theory of soliton gas/soliton turbulence in integrable systems. In this paper we shall be using the KdV soliton gas as a simplest analytically accessible model yielding a major insight into the general properties of soliton gases in integrable systems.

Macroscopic dynamics of a Korteweg--de Vries (KdV) soliton gas are determined by the fundamental `\emph{microscopic}' properties of two-soliton interactions \cite{John}: ({\it i}) soliton collisions are elastic, \ie~the interaction does not change the soliton amplitudes (or, more precisely, the discrete spectrum levels in the associated linear spectral problem for the quantum-mechanical Schr\"odinger operator); ({\it ii}) after the interaction, each soliton acquires an additional phase shift; ({\it iii}) the total phase shift of a `\emph{trial}' soliton acquired during a certain time interval can be calculated as a sum of the `\emph{elementary}' phase shifts in pairwise collisions of this soliton with other solitons during this time interval. Thus, the macroscopic dynamics of a soliton gas are essentially determined by two-soliton interactions. This fact enabled \textsc{Zakharov} in 1971 to introduce the kinetic equation for a `\emph{diluted}', small-density, gas of solitons for the KdV equation. The generalisation of \textsc{Zakharov}'s equation to finite densities derived in \cite{El2003} has required the consideration of the thermodynamic-type limits for finite-gap potentials and the associated Whitham modulation equations \cite{Whitham1999, Flaschka1980}. A straightforward, physical derivation of the kinetic equation was made in \cite{El2005a}. We stress that kinetic description of soliton gas makes an emphasis on the particle-like nature of solitons. At the same time, solitons represent nonlinear coherent wave structures so the total random nonlinear wave field associated with a soliton gas can be naturally interpreted as soliton turbulence \cite{El2001}. The effect of two-soliton collisions on the properties of the statistical moments of soliton turbulence was studied in \cite{Pelinovsky2013}. An effective method for the numerical computation of soliton gas was developed in \cite{Dutykh2013a} and was applied to the numerical modelling of soliton gas in the KdV and the KdV--BBM (Benjamin--Bona--Mahoni) equations in \cite{Dutykh2014d}.

The kinetic equation for solitons derived in \cite{El2003, El2005a} was shown in \cite{El2011a, Pavlov2012} to possess some remarkable mathematical properties. In particular, it was shown that it has an infinite number of integrable hydrodynamic reductions which is a strong indication of integrability of the full kinetic equation. Integrability of hydrodynamic reductions opens a broad perspective for obtaining various exact solutions to the kinetic equation. However, the quantitative confirmation of the relevance of the predictions of the soliton kinetic theory to the actual macroscopic dynamics of soliton gases still remains an open problem. Indeed, as was already mentioned, the formal derivation of the kinetic equation involves certain singular limiting  transition of the thermodynamic type. The mathematical conditions required for this transition are not necessarily applicable to physically (or even numerically) accessible soliton systems. This is why it is vitally important to have a direct numerical confirmation of the validity of the kinetic equation. The main goal of the pesent paper is thus to test the relevance of the soliton gas kinetics to the actual `\emph{particle dynamics}' of incoherent soliton ensembles. With this aim in view we compare some model exact solutions of the kinetic equation with the results of the high accuracy direct numerical simulations of the KdV soliton gas.

The paper is organised as follows. Section~\ref{sec:kin} provides a brief account on the kinetic equation for solitons. In Section~\ref{sec:hydro} hydrodynamic reductions of the kinetic equation are considered for one- and two-component soliton gases and two model problems are considered: the propagation of a trial soliton through a one-complonent cold soliton gas and the collision of two cold gases (the soliton gas shock tube problem). Section~\ref{sec:num} is devoted to the direct numerical modelling of the problems considered in Section~\ref{sec:kin} and detailed comparisons of the analytical and numerical solutions. The main conclusions of this study are outlined in Section~\ref{sec:concl}.


\section{Kinetic equation for a soliton gas}
\label{sec:kin}

We consider the KdV equation in the canonical form
\begin{equation}\label{KdV}
 u_t\ +\ 6\,uu_{x}\ +\ u_{xxx}\ =\ 0\,.
\end{equation}
We introduce soliton gas as an infinite collection of KdV solitons randomly distributed on the line with non-zero density. This intuitive definition lacks precision but it is sufficient for the purposes of this paper. A mathematically consistent definition of a soliton gas as the thermodynamic limit of finite-gap potentials will be briefly described below (see \cite{El2001} for details).

Let each soliton in the gas be `\emph{labeled}' by the spectral parameter $\eta_i\ \geq\ 0$ so that $\lambda_i\ =\ -\eta_i^2$ is the corresponding discrete eigenvalue in the spectral problem for the linear Schr\"odinger operator associated with \eqref{KdV} in the inverse scattering transform (IST) formalism. We assume that the discrete values $\eta_i$ are distributed with certain density on some finite interval, say $[0, 1]$ and replace $\eta_i$ by the continuous variable $\eta\ \in\ [0,1]$ (see the details in \cite{El2001, El2003, El2011a}). As is well known (see \eg \cite{John}) the amplitude of an \emph{isolated} KdV soliton with the spectral parameter $\eta$ is $a\ =\ 2\eta^2$ and its speed is $S\ =\ 4\eta^2$. It is clear that in the soliton gas, the mean speed of the soliton with the same parameter $\eta$ will differ from $4\eta^2$. Indeed, due to the pairwise collisions with other solitons (each leading to a `\emph{phase-shift}') the distance covered by this `\emph{trial}' soliton over some time interval $\Delta t\ \gg\ 1$ will be different from $4 \eta^2 \Delta t$. This intuitive reasoning was used in the original \textsc{Zakharov} paper \cite{Zakharov1971} for the derivation of the \emph{approximate} kinetic equation describing macroscopic dynamics of a `\emph{rarefied soliton gas}'.

The full, non-perturbative equation for a `\emph{dense}' gas of KdV solitons was derived in \cite{El2003} by considering a singular, thermodynamic type limiting transition for the modulation Whitham equations describing slow evolution of the finite-gap solutions of the KdV equation. The key property of the thermodynamic limit is the special spectral band-gap distribution (scaling) that preserves  finiteness of the \emph{integrated density of states} \cite{Lifshitz1988, Johnson1982} in the infinite-band limit. It was shown in \cite{El2001, El2003} that in the thermodynamic limit the integrated density of states yields the spectral measure $f(\eta)\,\ud\eta$ of the soliton gas. In a spatially inhomogeneous soliton gas $f(\eta) \equiv f(\eta; x,t)$ so that $f(\eta_0; x,t)\,\ud\eta\,\ud x$ is the number of solitons with the spectral parameter $ \eta \in (\eta_0,\ \eta_0 + \ud\eta)$ and located in the spatial interval $(x,\ x + \ud x)$ at the moment $t$. 
The integral
\begin{equation}\label{dens}
  \kappa(x,t)\ =\ \int_0^1 f(\eta, x,t)\;\ud\eta
\end{equation}
is the total physical (as opposed to spectral) density of the soliton gas, i.e. the number of solitons per unit length.

For the KdV equation \eqref{KdV} the evolution of the spectral density $f(\eta, x,t)$ is described by the integro-differential kinetic equation \cite{El2003}
\begin{equation}\label{kin1}
  f_t\ +\ [\,sf\,]_x\ =\ 0\,,
\end{equation}
\begin{equation}\label{s1}
  s(\eta)\ =\ 4\eta^2\ +\ \frac{1}{\eta}\;\int \limits^1_0 \ln\left|\frac{\eta + \mu}{\eta-\mu}\right|f(\mu)[s(\eta)-s(\mu)]\;\ud\mu\,,
\end{equation}
where we use the shorthand notations $f (\eta) \equiv f(\eta, x,t)$ and $s(\eta) \equiv s(\eta, x,t)$, the latter having the meaning of the mean soliton gas
velocity (or you can view it as the velocity of a `\emph{trial}' soliton with the spectral parameter $\eta$ placed in the soliton gas characterised by the distribution function $f(\mu)$). \textsc{Zakharov}'s approximate kinetic equation for a rarified soliton gas \cite{Zakharov1971} is obtained from \eqref{kin1}, \eqref{s1} by assuming $\kappa \ll 1$ and retaining only the first order correction in \eqref{s1}. It is important to stress that the typical scale of variations of $x$ and $ t$ in the kinetic equation \eqref{kin1} is much larger than in the KdV equation \eqref{KdV}, governing the primitive, `\emph{microscopic}' evolution.

It was noted in \cite{El2005a} that the formal procedure of the thermodynamic limit of the Whitham equations can be viewed as the justification of a simple straightforward derivation of the kinetic equations for solitons using the original \textsc{Zakharov}'s reasoning \cite{Zakharov1971} based on the phase shift expressions for two-soliton collisions. The key difference is that one should use the full, unapproximated mean velocity function $s(\eta)$ (rather than unperturbed free soliton velocity $4\eta^2$) in the calculation of the correction to the velocity of free soliton  which leads to a self-consistent definition of the local mean velocity of solitons with the spectral parameter $\eta$.

In the kinetic description \eqref{kin1}, \eqref{s1} of a soliton gas the solitons are viewed as particles moving with certain speeds and interacting with each other according to the nonocal closure equation \eqref{s1}. On the other hand, one is also interested in the nonlinear wave field $u(x,t)$ associated with the soliton gas dynamics (an integrable soliton turbulence). As is well known (see \eg \cite{Monin2007, Nazarenko2011}) a turbulent wave field is usually characterised by the moments $\langle u^n \rangle$ over the statistical ensemble which, due to ergodicity of soliton turbulence \cite{El2001}, can be computed as spatial averages $\overline{u^n} = \frac{1}{\Delta}\int ^\Delta_0 u^n(\tilde x, t)d \tilde x$, over a sufficiently large interval $1 \ll \Delta  \ll L$, where $L$ is the typical scale for $x$-variations in (\ref{kin1}).

It was shown in \cite{El2001, El2003, El2015} that the two first moments in the KdV soliton turbulence are calculated in terms of the spectral distribution function $f(\eta, x,t)$ as
\begin{equation}\label{mean}
  \overline{u}(x,t)\ =\ 4\int_{0}^{1}\eta f(\eta, x,t )\;\ud\eta \,, \qquad \overline{u^{2}}(x,t)\ =\ \dfrac{16}{3}\int_{0}^{1}\eta ^{3}f(\eta,x,t)\;\ud\eta \,.
\end{equation}

A fundamental restriction imposed on the distribution function $f(\eta)$ follows from non-negativity of the variance
\begin{equation}\label{var}
  \mathcal{A}^2\ :=\ \overline{u^2}\ -\ \overline{u}^2\ \geq\ 0.
\end{equation}
The consequences of this restriction have been explored in \cite{El2015}. Here it will inform the choice of the soliton gas parameters for the numerical modelling.


\section{Hydrodynamic reductions and exact solutions}
\label{sec:hydro}

To get a better insight into the properties of the soliton gas dynamics we consider the hydrodynamic reductions of the kinetic equation \eqref{kin1}, \eqref{s1}. Such hydrodynamic reductions enable one to derive some simple, physically interesting exact solutions which could then be compared with the results of direct numerical modelling of the KdV equation.

The family of the simplest $N$-component ($N$-beam) `\emph{cold gas}' reductions is selected by the delta-function multiflow ansatz, well known in plasma physics \cite{Silin1971}:
\begin{equation}\label{delta}
  f(\eta, x,t)\ =\ \sum \limits_{i=1}^{N}\, f_i(x,t)\,\delta(\eta\ -\ \eta_i) \, .
\end{equation}
Physically, the $i$-th complonent of the soliton gas described by the distribution \eqref{delta} consists of an infinite sequence of nearly identical solitons having the spectral parameter $\eta$ distributed in a narrow $\varepsilon$-vicinitty of $\eta = \eta_i$ such that $\varepsilon/\eta_i \ll 1$, and distributed by Poisson on $x \in (-\infty, \infty)$ with the density $f_i(x,t)$ which can slowly vary in space in time \cite{Pavlov2012, El2015}.

The hydrodynamic reductions obtained by \eqref{delta} for arbitrary $N$ have been thoroughly analysed in \cite{El2011a, Pavlov2012} where their integrability in the sense of the generalised hodograph transform was proven. Here we shall be mostly looking at the case of a two-component gas yielding the simplest nontrivial results that can be verified numerically. We consider two model problems: ({\it i}) the propagation of a `\emph{trial}' soliton through a one-component `\emph{cold}' soliton gas and ({\it ii}) collision of two one-component soliton gases --- the shock tube problem.


\subsection{Propagation of a trial soliton through one-component soliton gas}
\label{sec:31}

We consider a `\emph{trial}' soliton with the spectral parameter $\eta = \eta_1$ moving through a one-component soliton gas with the distribution function 
\begin{equation}\label{cold1}
  f(\eta;\, x,t)\ =\ f_0(x,t)\;\delta(\eta\ -\ \eta_0),
\end{equation}
where the density $f_0(x,t)$ is found by the substitution of the distribution \eqref{cold1} into \eqref{kin1}, \eqref{s1}, yielding the linear transport equation
\begin{equation}\label{lin}
  \pd{f_0}{t}\ +\ 4\eta_0^2\;\pd{f_0}{x}\ =\ 0\,,
\end{equation}
Equation \eqref{lin} describes a trivial translation of the initial distribution function with the constant speed $s(\eta_0)=4 \eta_0^2$, \ie~$f_0(x,t) = F(x - 4\eta_0^2 t)$, where $F(x) \equiv f_0(x,0)$ is the initial distribution. Substituting the distribution function \eqref{cold1} into the expressions \eqref{mean} for the moments we obtain from \eqref{var} the restriction for the soliton gas density \cite{El2015}
\begin{equation}\label{crit_dens}
  f_0\ \leq\ \frac{\eta_0}{3}.
\end{equation}

The mean velocity of the trial soliton $s(\eta_1; x,t)$ can then be found from formula \eqref{s1},
\begin{equation}\label{s11}
  s_1(x,t)\ =\ 4\eta_1^2\ +\ \frac{1}{\eta_1}\;\ln\left|\frac{\eta_1\ +\ \eta_0}{\eta_1\ -\ \eta_0}\right|\,f_0(x,t)\;\bigl[\,s_1(x,t)\ -\ 4\eta_0^2\,\bigr]\,,
\end{equation}
where $s_1(x,t)\ \equiv\ s(\eta_1; x,t$)). Expressing $s_1$ from \eqref{s11} we obtain
\begin{equation}\label{s111}
  s_1\ =\ 4\;\frac{\eta_1^2\ -\ \alpha\eta_0^2 f_0(x,t)}{1\ -\ \alpha f_0(x,t)}\, , \qquad \alpha f_0(x,t)\ \ne\ 1.
\end{equation}
Here
\begin{equation}\label{shift}
  \alpha\ =\ \frac{1}{\eta_1}\,\ln\left|\frac{\eta_1\ +\ \eta_0}{\eta_1\ -\ \eta_0}\right|\ >\ 0
\end{equation}
is the classical phase-shift expression for the two-soliton collision \cite{John}.

\subsection{Soliton gas shock tube problem}

We now consider a two-component soliton gas by introducing the distribution function in the form 
\begin{equation}\label{cold2}
  f(\eta, x,t)\ =\ f_1(x,t)\;\delta(\eta\ -\ \eta_1)\ +\ f_2(x,t)\;\delta(\eta\ -\ \eta_2)\,,
\end{equation}
where 
\begin{equation}\label{pm}
  \eta_{1,2}\ >\ 0\,, \qquad \eta_1\ \ne\ \eta_2\,, \qquad f_{1,2}\ \geq\ 0\,.
\end{equation}

Substitution of \eqref{cold2} into \eqref{kin1}, \eqref{s1} leads to the system of two conservation laws
\begin{equation}\label{red1}
  \partial_t f_1\ +\ \partial_x[\,f_1 s_1\,]\ =\ 0\,, \qquad \partial_t f_2\ +\ \partial_x [\,f_2 s_2\,]\ =\ 0,
\end{equation}
\begin{equation}\label{sf}
  s_1\ =\ 4\eta_1^2\ +\ \alpha_1(s_1\ -\ s_2) f_2 \,, \qquad
  s_2\ =\ 4\eta_2^2\ +\ \alpha_2(s_2\ -\ s_1) f_1\,,
\end{equation}
where $s_{1,2}(x,t)\ \equiv\ s(\eta_{1,2}, x,t)$ and
\begin{equation}\label{epsilon}
  \alpha_{1,2}\ =\ \frac{1}{\eta_{1,2}}\,\log\left|\frac{\eta_1\ +\ \eta_2}{\eta_1\ -\ \eta_2}\right|\ >\ 0 \,.
\end{equation}
It follows from \eqref{pm}, \eqref{sf}, \eqref{epsilon} that if $s_1\ >\ 4\eta_1^2$ then $s_1\ >\ s_2$ and $s_2\ <\ 4\eta_2^2$. Similarly, if $s_1\ <\ 4\eta_1^2$ then $s_1\ <\ s_2$ and $s_2\ >\ 4\eta_2^2$.

We note that, in the degenerate case $\eta_1 = \eta_2 \equiv \eta_0$ the ansatz \eqref{cold2} reduces to the one-component distribution \eqref{cold1} with $f_0 = f_1 + f_2$ as expected. Expressing $s_{1,2}$ in terms of $f_{1,2}$ from \eqref{sf} we obtain
\begin{equation}\label{s120}
  s_1\ =\ 4\eta_1^2\ +\ \frac{4(\eta_1^2\ -\ \eta_2^2)\alpha_2 f_2}{1\ -\ \alpha_1 f_1\ -\ \alpha_2 f_2}, \qquad
  s_2\ =\ 4\eta_2^2\ -\ \frac{4(\eta_1^2\ -\ \eta_2^2)\alpha_1 f_1}{1\ -\ \alpha_1 f_1\ -\ \alpha_2 f_2}
\end{equation}
provided $f_1 \alpha_1\ +\ f_2 \alpha_2\ \ne\ 1$. Note that by setting $f_1 \equiv 0$ we recover the expression \eqref{s111} for the speed of the trial soliton $s_1$ (to establish the correspondence with \eqref{s111} the index `$2$' in the first expression \eqref{s120} should be replaced with $0$, also $\alpha_2$ becomes $\alpha$). The density of the two-component soliton gas \eqref{dens} is
\begin{equation*}
  \kappa\ =\ f_1\ +\ f_2.
\end{equation*}

It is not difficult to show that equations \eqref{red1}, \eqref{sf} assume Riemann invariant form in variables $s_1, s_2$ \cite{El2005a},
\begin{equation}\label{Riemann}
  \partial_t s_1\ +\ s_2 \partial_x\, s_1\ =\ 0\,, \qquad 
  \partial_t s_2\ +\ s_1\partial_x\, s_2\ =\ 0\,.
\end{equation}
System \eqref{Riemann} is linearly degenerate \cite{Rozhdestvenskiy1978} which implies: ({\it i}) the absence of the nonlinear wave-breaking effects in a two-component soliton gas (see \cite{El2011a} for the relevant account of the properties of linearly degenerate hydrodynamic type systems) and ({\it ii}) unavailability of simple-wave solutions (indeed one can easily see that $s_2(s_1)$ implies that $s_{1,2}$ are constants). The component densities $f_{1,2}(x,t)$ are expressed in
terms of the velocities $s_{1,2}$ by the expressions derived from \eqref{sf}
\begin{equation}\label{f12}
  f_1\ =\ \frac{s_2\ -\ 4\eta_2^2}{\alpha_2(s_2\ -\ s_1)}\,, \qquad
  f_2\ =\ \frac{s_1\ -\ 4\eta_1^2}{\alpha_1(s_1\ -\ s_2)}\,.
\end{equation}

Using the distribution function \eqref{cold2} we obtain for the two first moments \eqref{mean} of the wave field in the two-component gas:
\begin{equation}\label{mom2c}
  \overline{u}\ =\ 4\;(\eta_1 f_1\ +\ \eta_2 f_2), \qquad 
  \overline{u^2}\ =\ \frac{16}{3}\;(\eta_1^3 f_1\ +\ \eta_2^3 f_2)\,.
\end{equation}
Thus, equations \eqref{Riemann}, \eqref{f12} completely define the evolution of the moments \eqref{mom2c} of the nonlinear wave field associated with the two-component soliton gas. The nonnegativity of the variance $\mathcal{A}^2$ defined in \eqref{var} implies the condition (\cf the condition \eqref{crit_dens} for the single-component gas):
\begin{equation}\label{var2}
  \eta_1 f_1\, \Bigl(\frac{\eta_1^2}{3}\ -\ \eta_1 f_1\ -\ \eta_2 f_2\Bigr)\ +\ \eta_2 f_2\, \Bigl(\frac{\eta_2^2}{3}\ -\ \eta_1 f_1\ -\ \eta_2 f_2\Bigr)\ \geq\ 0.
\end{equation}


We now consider the Riemann problem for the two-component soliton gas characterized by the spectral distribution function \eqref{cold2} corresponding to the shock tube problem: an initial contact dscontinuity separating gases of different density
\begin{equation}\label{RPf}
\left\{\begin{array}{ll}
  f_1 (x,0)\ =\ f_{10}, \quad f_2(x,0)\ =\ 0\,, \quad & x\ <\ 0, \\
  f_2 (x,0)\ =\ f_{20}, \quad f_1(x,0)\ =\ 0\,, & x\ >\ 0,
\end{array}
\right.
\end{equation}
where $f_{10}$, $f_{20}\ >\ 0$ are some constants satisfying $f_{i0}\ \leq\ \eta_i/3$ (see \eqref{crit_dens}). We also assume that $\eta_1\ >\ \eta_2$. Note that, unlike in the classical gas-dynamics shock tube problem, the initial  velocity of the soliton gases is not zero but is fully determined by the density distribution \eqref{RPf} via relations \eqref{s120},
\begin{equation}\label{RPs}
  \left\{\begin{array}{ll}
  s_1 (x,0)\ =\ 4\eta_1^2\ \equiv\ s_{10}, \quad s_2(x,0)\ =\ 4\eta_2^2\ -\ \dfrac{4(\eta_1^2\ -\ \eta_2^2)\alpha_1 \rho_{10}}{1\ -\ \alpha_1 \rho_{10}}, \quad & x\ <\ 0, \\
  s_1(x,0)\ =\ 4\eta_1^2\ +\ \dfrac{4(\eta_1^2\ -\ \eta_2^2)\alpha_2\rho_{20}}{1\ -\ \alpha_2\rho_{20}}, \quad s_2(x,0)\ =\ 4 \eta_2^2\ \equiv\ s_{20}, & x\ >\ 0.
\end{array}
\right.
\end{equation}
We note that the values of $s_2$ for $x < 0$ and $s_1$ for $x > 0$ corresponds to the respective zero-density components (\cf \eqref{RPf}) and thus, are  fictitious initial parameters providing consistency with the equation to be solved.

\begin{figure}
  \centering
  \includegraphics[width=0.59\textwidth]{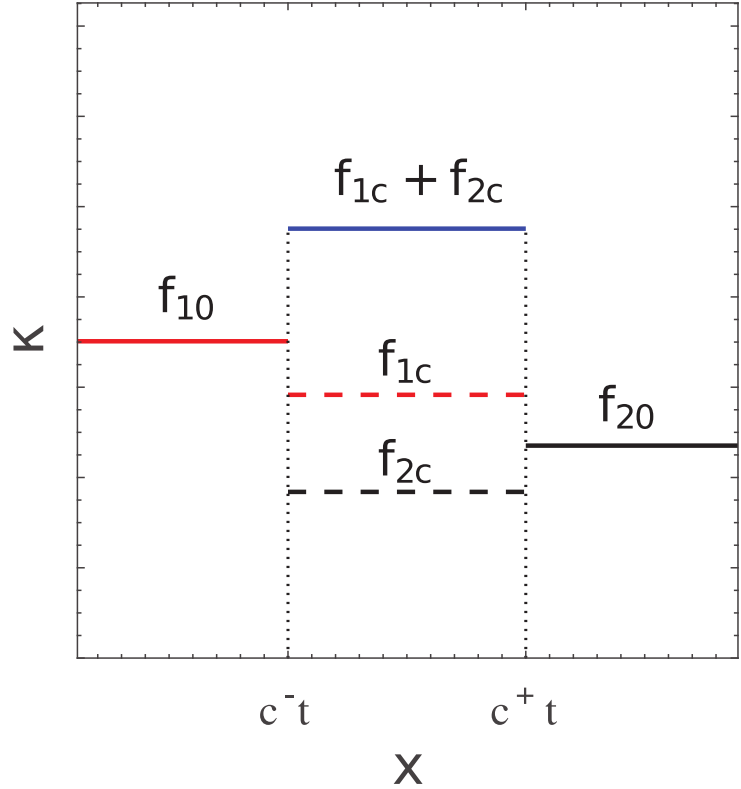}
  \caption{\small\em Weak solution $\kappa(x,t)$ of the soliton gas shock tube problem (solid lines). Dashed lines: densities of individual components in the interaction region.}
  \label{fig1}
\end{figure}

Since the governing equations \eqref{Riemann} are quasilinear the solution of the Riemann problem must depend on $x/t$ alone. Since due to linear degeneracy system \eqref{Riemann} does not have non-constant simple wave solutions one has to look for weak similarity solutions of the original conservation laws \eqref{red1}. The required solution  represents a combination of three constant states separated by two  contact discontinuities (see \cite{El2005a}). For the total density $\kappa\ =\ f_1\ +\ f_2$ we have
\begin{equation}\label{RPs}
  \kappa(x,t)\ =\ \left\{\begin{array}{ll}
    f_{10}, \quad & x\ <\ c^-t, \\
    f_{1c}\ +\ f_{2c}, \quad & c^-t\ <\ x\ <\ c^+t, \\
    f_{20}, \quad & x\ >\ c^+t.
\end{array}
\right.
\end{equation}
The schematic representation of this solution is shown in Figure~\ref{fig1}.

The values $f_{1c}$ and $f_{2c}$ as well as the velocities $c^{\pm}$ of the discontinuities are found from the Rankine--Hugoniot jump conditions 
\begin{equation}
\begin{array}{l}
  -c^-(f_{10}\ -\ f_{1c})\ +\ (f_{10}s_{10}\ -\ f_{1c}s_{1c})\ =\ 0\,, \\
  -c^-(0\ -\ f_{2c})\ +\ (0\ -\ f_{2c} s_{2c})\ =\ 0\,,
\end{array}
\label{j1}
\end{equation}
\begin{equation}
\begin{array}{l}
  -c^+(f_{1c}\ -\ 0)\ +\ (f_{1c} s_{1c}\ -\ 0)\ =\ 0 \,, \\
  -c^+(f_{2c}\ -\ f_{20})\ +\ (f_{2c}s_{2c}\ -\ f_{20}s_{20})\ =\ 0 \,.
\end{array}
\label{j2}
\end{equation}
Here $c^-$ and $c^+$ are the velocities of the left and right discontinuities respectively, and $f_{1c}$, $f_{2c}$ and $s_{1c}$, $s_{2c}$ are the densities and velocities of the soliton gas components in the interaction region $x\ \in\ [c^-t, \ c^+ t]$. The velocities $s_{1c}$ and $s_{2c}$ are expressed in terms of $f_{1c}$, $f_{2c}$ by relations \eqref{s120}.

Solving \eqref{j1} and \eqref{j2} we obtain:
\begin{equation}\label{fc}
  f_{1c}\ =\ \frac{f_{10}(1\ -\ \alpha_2 f_{20})}{1\ -\ \alpha_1 \alpha_2 f_{10} f_{20}}\,, \quad f_{2c}\ =\ \frac{f_{20}(1\ -\ \alpha_1 f_{10})}{1\ -\ \alpha_1 \alpha_2 f_{10} f_{20}}\,.
\end{equation}
The speeds $c^{\pm}$ of the boundaries of the interaction region are given by
\begin{equation*}
  c^-\ =\ 4\eta_2^2\ -\ \frac{4(\eta_1^2\ -\ \eta_2^2)\eta_1 f_{1c}}{1\ -\ \alpha_1 f_{1c}\ -\ \alpha_2 f_{2c}}\,, \quad 
  c^+\ =\ 4 \eta_1^2\ +\ \frac{4(\eta_1^2\ -\ \eta_2^2) \alpha_2 f_{2c}}{1\ -\ \alpha_1 f_{1c}\ -\ \alpha_2 f_{2c}}\,.
\end{equation*} 

The expanding interaction region in the soliton gas shock tube problem  can be viewed as an \emph{incoherent dispersive shock wave}, a stochastic counterpart of the traditional, coherent dispersive shock wave (DSW) forming due to a dispersive regularisation of the Riemann initial data in the KdV equation \cite{Gurevich1974a} (we need to make a clear disctinction between the studied here incoherent DSWs, which are generated in the Riemann problems for soliton gases and the incoherent DSWs recently observed in Fourier spectra evolution of random waves \cite{Garnier2013}). In contrast with the coherent DSWs, the incoherent DSW generated in the collision of two soliton gases does not have a disctinct structure of a slowly modulated wavetrain but is characterised by the increased itensity of fluctuations $\mathcal{A}^2$  \eqref{var} of the random nonlinear wave field, compared to the values of $\mathcal{A}^2$ in the colliding soliton gases at $t = 0$. It is not difficult to show that the density of the two-component soliton gas in the incoherent DSW region $\kappa_c\ =\ f_{1c} + f_{2c}\ >\ f_{10}, f_{20}$ but $\kappa_c\ <\ f_{10} + f_{20}$ (see Figure~\ref{fig1}).

In conclusion of this section we note that the above results for a two-component gas cannot be \emph{in principle} derived from the approximate kinetic equation obtained in \cite{Zakharov1971} for a rarefied gas. Indeed, it was shown in \cite{Zakharov1971} that the kinetic equation for a rarified soliton gas prescribes linear instability for a two-component gas so that any density perturbation would grow without bound making the underlying equation inapplicable. The full kinetic equation used here yields hyperbolic hydrodynamic reduction \eqref{Riemann} for a two-component gas implying stability for the full range of admissible densities. This conclusion will be confirmed by direct numerical simulations  in the next section.


\section{Numerical experiments}
\label{sec:num}

In this section we perform  direct numerical simulations of the KdV soliton gas  and compare the numerical solutions with the corresponding solutions of the kinetic equation obtained in Section~\ref{sec:hydro}.

\subsection{Numerical method}

In order to solve numerically the KdV equation we employ the standard pseudo-spectral Fourier collocation technique \cite{Trefethen2000, Boyd2000}. This method is briefly explained below.

\begin{figure}
  \centering
  \includegraphics[width=0.89\textwidth]{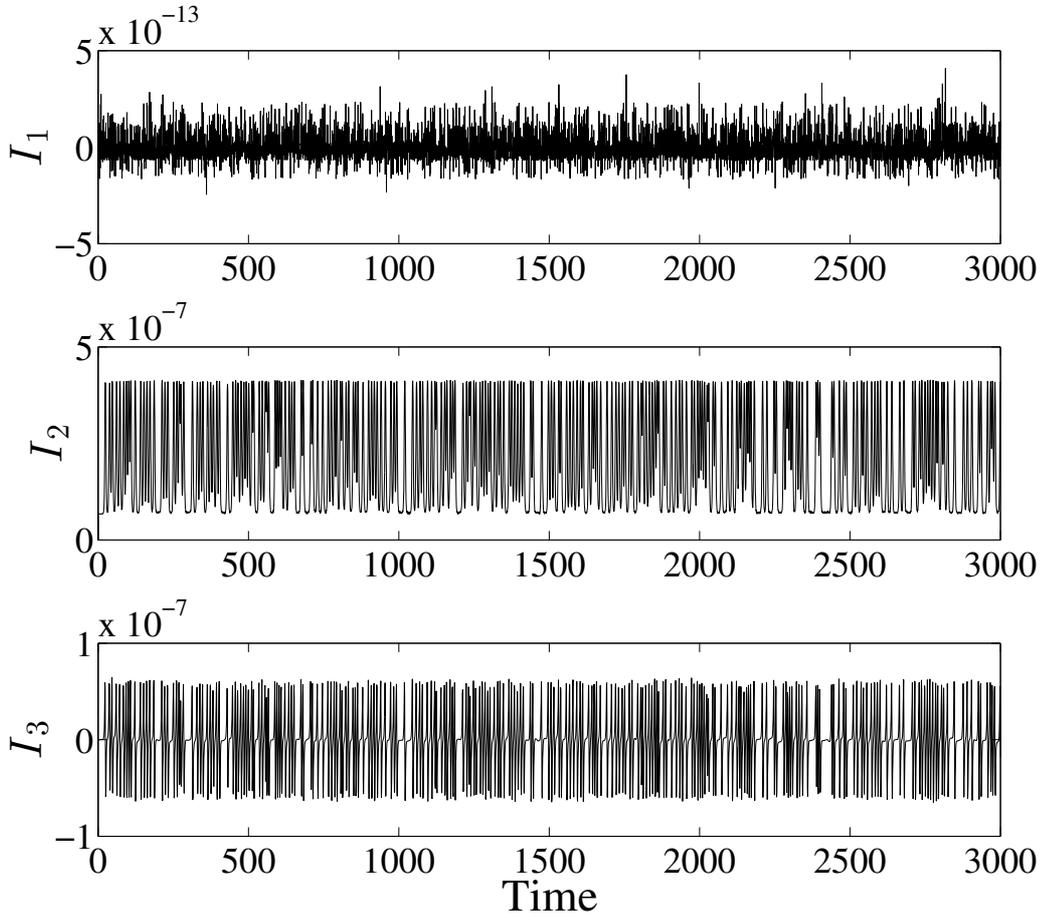}
  \caption{\small\em Evolution of the error in the three first invariants of the KdV equation.}
  \label{fig2}
\end{figure}

Denote by $\uh(k,t) = \F\{u\}$ the Fourier transform of $u(x,t)$ in $x$, where $k$ is the wavenumber. Then, by Fourier-transforming the KdV equation \eqref{KdV} yields
\begin{equation}\label{eq:sp}
  \uh_t\ -\ \ui k^3\uh\ =\ -3 \ui k\widehat{(u^2)}.
\end{equation}
The most computationally efficient way consists in computing the spatial derivatives in spectral space while the nonlinear product is computed in real space and de-aliased using the classical $3/2$th rule. The overall implementation is very efficient thanks to the FFT algorithm. In order to improve the time-stepping we will use the so-called integrating factor technique. This consists of the exact integration of the linear terms of \eqref{eq:sp}, viz. 
\begin{equation}\label{eq:spectral}
  \hat{v}_t\ =\ \ue^{(t-t_0)\L}\cdot\Nn\Bigl\{\ue^{-(t-t_0)\L}\cdot\hat{v}\Bigr\},
  \qquad 
  \hat{v}(t)\ \equiv\ \ue^{(t-t_0)\L}\cdot\uh(t), \qquad 
  \hat{v}(t_0)\ =\ \uh(t_0),
\end{equation}
where the linear and nonlinear operators $\L$ and $\N$ are defined through their symbols as
\begin{equation*}
  \L\ :=\ -\ui k^3, \qquad \Nn\ :=\ -3 \ui k\widehat{(u^2)}.
\end{equation*}
This allows to increase substantially the accuracy and the stability region of the time marching scheme (see, for example \cite{Trefethen2000}). Finally, the resulting system of ODEs is discretized in time by the Verner's embedded adaptive 9(8) Runge--Kutta scheme \cite{Verner1978}. The time step is adapted automatically according to the H211b digital filter approach \cite{Soderlind2003}. In order to estimate the accuracy of the numerical solution, one can follow the values of quantities which are conserved during the evolution. For the KdV equation we check the values of the first three invariants during all numerical computations. A typical evolution of these invariants is represented in Figure~\ref{fig2}.

\subsection{Test 1: Propagation of a trial soliton through a one-component gas}

We now present the results of the numerical simulation of the  propagation of a trial soliton with given spectral parameter $\eta\ =\ \eta_1$ through the one-component soliton gas with $\eta\ =\ \eta_0$ to compare with the theoretical results of Section~\ref{sec:31}. The value for the comparison is the mean (\ie~averaged over a large interval) velocity of the trial soliton is given by formula (\ref{s111}) in which $f_0\ =\ \const$.

\begin{figure}
  \centering
  \includegraphics[width=0.99\textwidth]{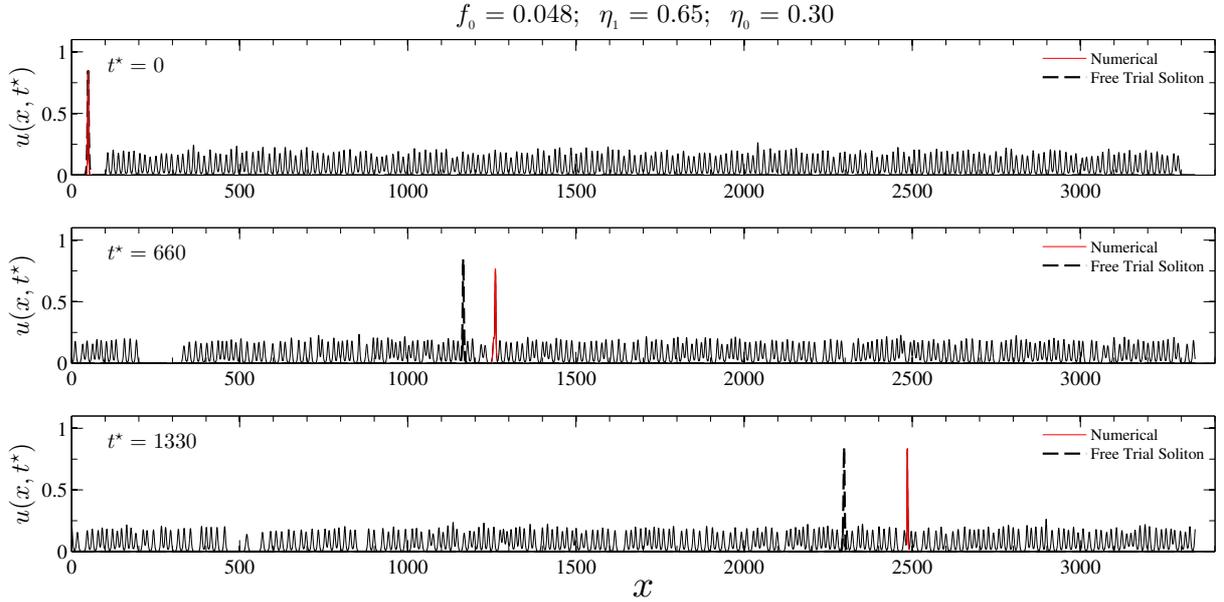}
  \caption{\small\em Comparison for the propagation of a free soliton with $\eta_1\ =\ 0.65$ in a void (black dashed line) with the propagation of the trial soliton with the same spectral parameter (red solid line) through a soliton gas, with the dominant spectral component $\eta_0\ =\ 0.3$ and density $f_0\ =\ 0.048$. One can see that the trial soliton gets accelerated due to the interactions with smaller solitons in the gas.}
\label{fig3}
\end{figure}

In the simulations, the initial condition for the one-component KdV soliton gas is composed of finite but sufficiently large number $M$ of solitons (in our experiments $M = 200$) with the random amplitude $a = 2\eta^2$ chosen from the  normal distribution of the spectrum $\eta$ with mean $\eta_0$ and fixed standard deviation $\sigma = 2\times 10^{-2}$, separated by a space lag $\Delta_0$ whose value is directly related to the gas density:
\begin{equation*}
  w(x,0)\ =\ \sum \limits_{i=1}^{M}\, 2\eta_i^2\;\hbox{sech}^2\bigl(\,\eta_i\,[x\ -\ (\ell\ +\ i\Delta_0\ +\ \epsilon_i)]\,\bigr).
\end{equation*}

The parameters $\ell$ and $\epsilon_i$ are respectively the starting point of the random lattice and a random (uniform) perturbation to the $i$-th soliton position, taken in the interval $\epsilon_i \in [-1, 1]$. So that, in order to increase (or decrease) the density $\kappa_0$, it is only required to change the value of the space lag $\Delta_0$.

With an added trial $\eta_1$-soliton the initial-boundary conditions for the KdV equation \eqref{KdV} assume the form 
\begin{align*}
  u(x,\, 0)\ &=\ 2\eta_1^2\;\hbox{sech}^2(\eta_1 x)\ +\ w(x,0), \\ 
  u(x+2L,\, t)\ &=\ u(x,t).
\end{align*}

\begin{figure}
  \centering
  \includegraphics[width=0.99\textwidth]{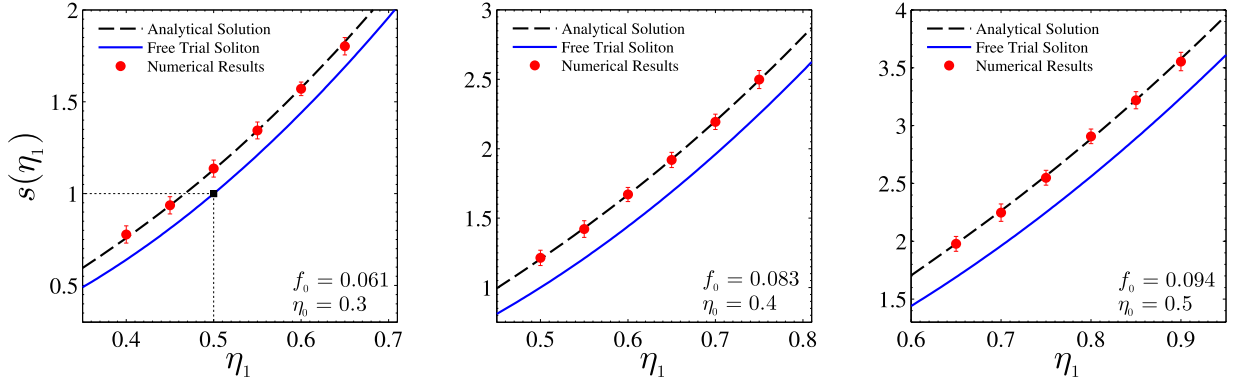}
  \caption{\small\em Comparison of the kinetic theory prediction \eqref{s11} for the average speed of a large trial soliton propagating through a one-component soliton gas with the results of direct numerical simulations of the KdV soliton gas.}
  \label{fig4}
\end{figure}

The snapshots of the trial soliton evolution are shown in Figure~\ref{fig3}.  One can see that the trial soliton undergoes a noticeable acceleration as predicted by the theory. The quantitative comparisons of the numerically found values for the averaged speed of the trial soliton with the formula \eqref{s11} are shown in Figure~\ref{fig4} for three different sets of parameters of the soliton gas.

The comparisons show an excellent agreement between the results of direct numerical simulations and the predictions of the kinetic theory. In all simulations the condition \eqref{crit_dens} for the soliton gas density is satisfied.

\subsection{Test 2: Soliton gas shock tube problem}

Following the strategy proposed in the previous section, we build the initial condition as a superposition of two distinct populations of solitons separated at $t^\star = 0$ by an empty gap, so that
\begin{equation*}
  u(x,0)\ =\ w_1(x,0)\ +\ w_2(x,0),
\end{equation*}
where
\begin{equation}\label{2comp}
  w_\alpha(x,0)\ =\ \sum\limits_{i=1}^{N}\,2\eta_{\alpha,i}^2\;\hbox{sech}^2 \bigl(\eta_{\alpha,i}\,[x\ -\ (\ell_\alpha\ +\ i\Delta_\alpha\ +\ \epsilon_{\alpha,i})]\bigr)\,,
\end{equation}
with $\alpha$ represent the single gas component ($\alpha = 1,2$). As in the pevious case the amplitudes of the two gas components are Gaussian random values distributed with the means $\eta_1$ and $\eta_2$ and standard deviations $\sigma_1\ =\ 10^{-4}$ and $\sigma_2\ =\ 2 \times 10^{-2}$ respectively. Again, the respective densities $f_{10}$ and $f_{20}$ can be easily changed by tuning the parameters $\Delta_1$ and $\Delta_2$ in equation \eqref{2comp}.

\begin{figure}
  \centering
  \includegraphics[width=0.99\textwidth]{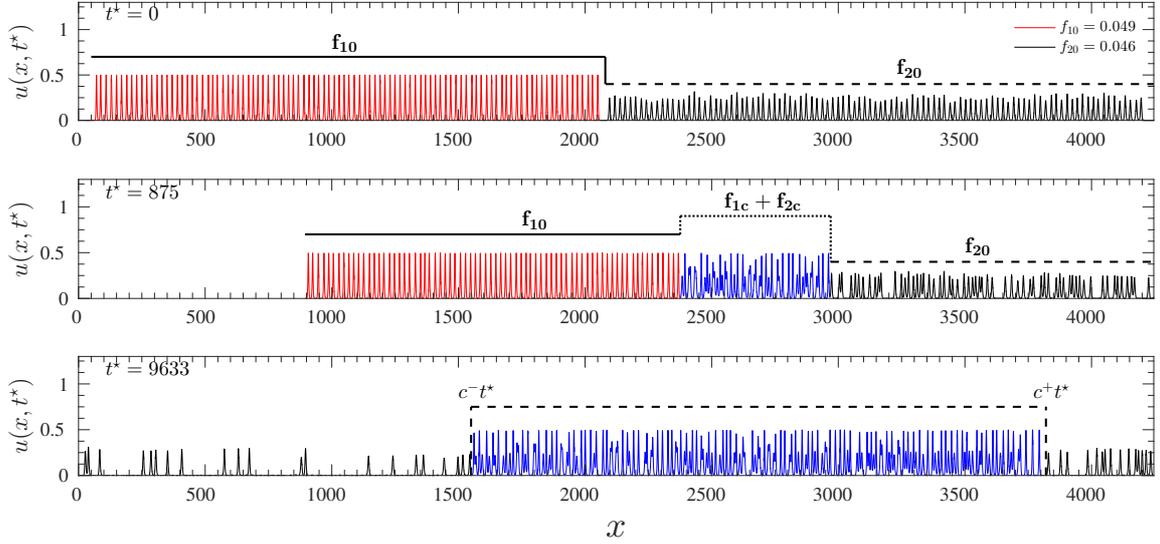}
  \caption{\small\em Soliton gas shock tube problem: numerical solution of the KdV equation. The expanding incoherent DSW region forming due to the interaction of two cold soliton gases is shown in blue.}
  \label{fig5}
\end{figure}

The numerical solution of the KdV equation with initial condition \eqref{2comp} is presented in Figure~\ref{fig5}. We now perform the comparison of the parameters of this numerical solution with the weak analytical solution of the soliton gas shock tube problem obtained in Section~\ref{sec:31}. Specifically, we are interested in comparing the total density of solitons $\kappa\ =\ f_{1c}\ +\ f_{2c}$ in the interaction (incoherent DSW) region and in the speeds $c^{\pm}$ of its edges.

\begin{figure}
  \centering
  \includegraphics[width=0.99\textwidth]{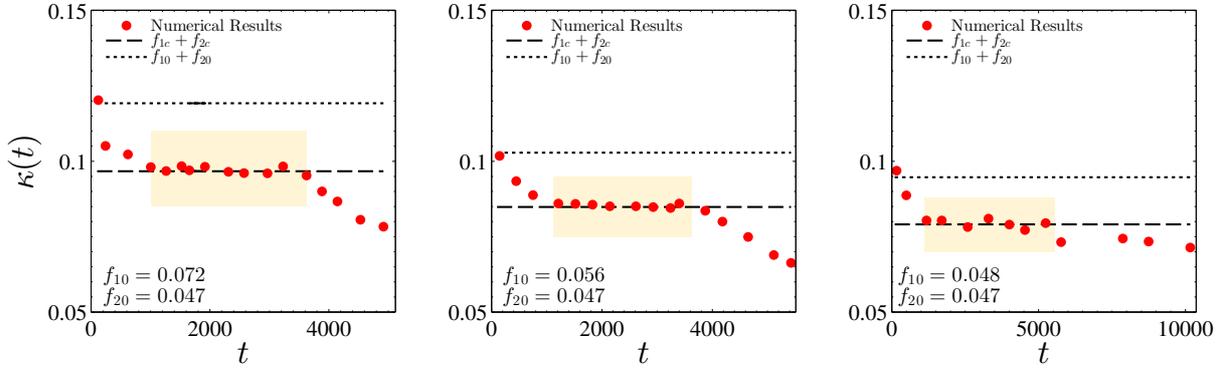}
  \caption{\small\em Shock tube problem: total density in the interaction region as a function of time. The highlighted regions correspond to the equilibrium state.}
  \label{fig6}
\end{figure}

The first observation is that the incoherent DSW forming due to the interaction of two cold soliton gases is stable in agreement with the hyperbolic nature of the two-component hydrodynamic reduction \eqref{Riemann} of the full kinetic equation. The comparisons for the total density as the function of time in the interaction region is presented in Figure~\ref{fig6}. One can see three distinct regions in the presented numerical plots. The value of the total density is initially equal to the sum $f_{10}\ +\ f_{20}$ of the component densities and then decreases through the equilibration process to the stationary value $\kappa_c$ (highlighted in all threee plots) which is in excellent agreement with the predictions of the theory based on the weak solution \eqref{RPs} of the two-component hydrodynamic reduction of the kinetic equation. The subsequent decrease of the density seen in the numerical plots is due to an inherent restriction of the numerical experiment involving finite number of solitons so the interaction region is sustained only for a finite interval of time.

\begin{figure}
  \centering
  \includegraphics[width=0.99\textwidth]{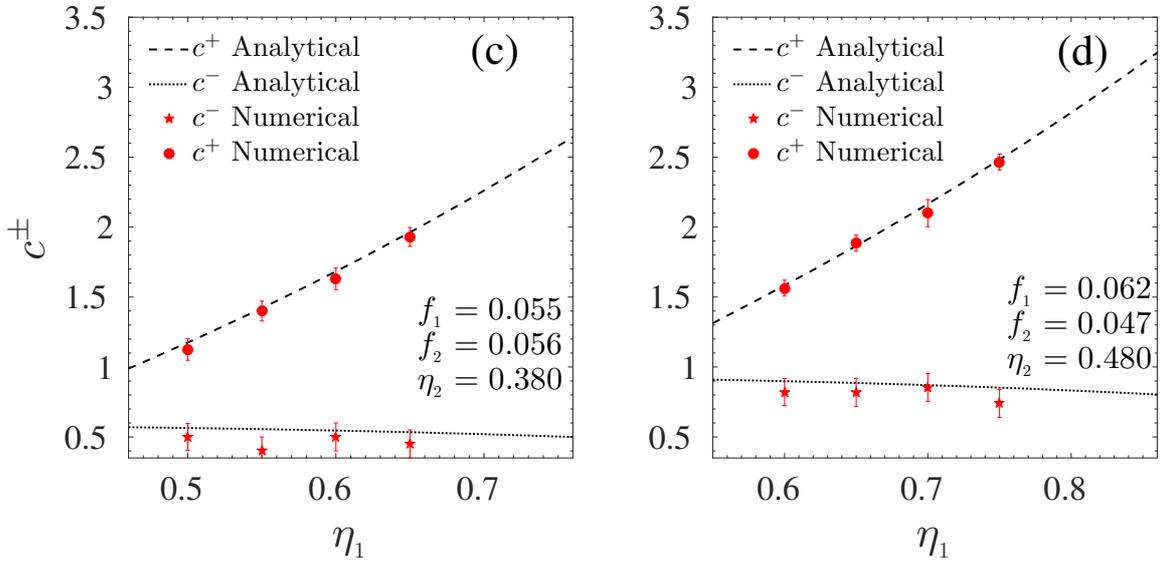}
  \caption{\small\em Comparisons for the shock tube problem: the speeds $c^\pm$ of the edges of the interaction region.}
  \label{fig7}
\end{figure}

The comparisons for the velocities of the edges of the interaction region (incoherent DSW) is presented in Figure~\ref{fig7} and demonstrate excellent agreement between the analytical and numerical results.

\begin{figure}
  \centering
  \subfigure[]{\includegraphics[width=0.49\textwidth]{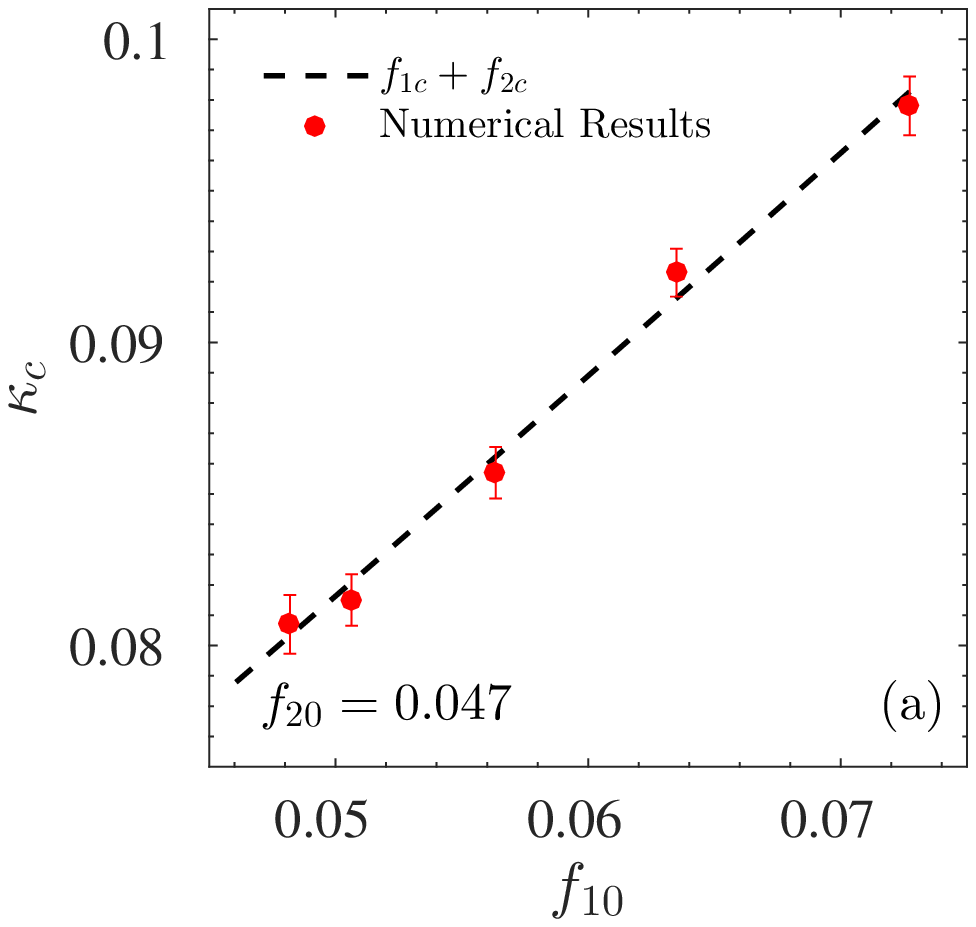}}
  \subfigure[]{\includegraphics[width=0.49\textwidth]{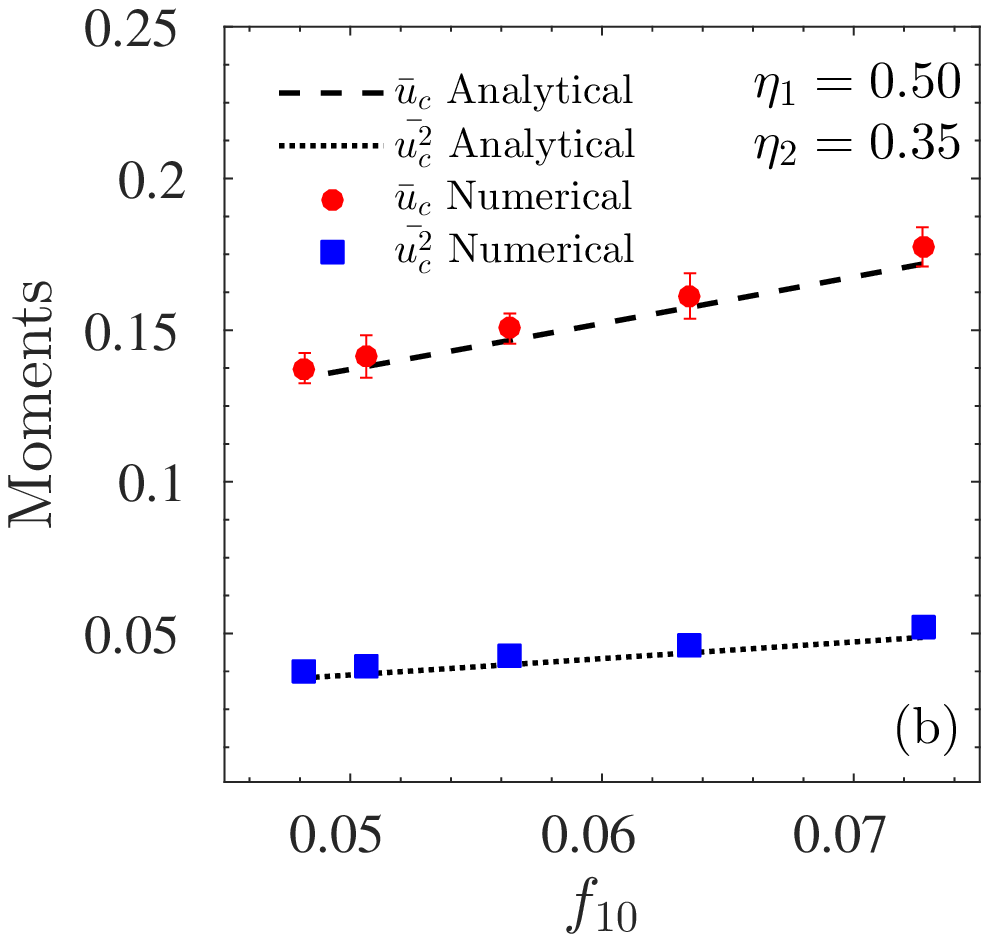}}
  \caption{\small\em Soliton gas shock tube problem: (a) the equlibrium total density $\kappa_c\ =\ f_{1c}\ +\ f_{2c}$ in the interaction region as a function of the density $f_{10}$ of the `left' gas. The density of the `right' gas $f_{20}\ =\ 0.047$. (b) the moments of the random wave field in the interaction region.}
  \label{fig8}
\end{figure}

We shall also compare the values of the two first moments \eqref{mom2c}, which for the interaction region assume the form 
\begin{equation}\label{mom2c1}
  (\overline u)_c\ =\ 4\;(\eta_1\, f_{1c}\ +\ \eta_2\, f_{2c}); \qquad 
  (\overline {u^2})_c\ =\ \frac{16}{3}\;(\eta_1^3\, f_{1c}\ +\ \eta_2^3\, f_{2c})\,.
\end{equation}
where $f_{1c}$ and $f_{2c}$ are determined in terms of the initial data by formulae \eqref{fc}. The results of the comparison are presented in Fig.~\ref{fig8}b. Again, the excellent agreement  is observed. One can also see that the condition \eqref{var2}, $\mathcal{A}_c^2\ =\ (\overline{u^2})_c\ -\ (\overline{u})_c^2\ >\ 0$, is satisfied.


\section{Conclusions}
\label{sec:concl}

We have undertaken a detailed comparison of the macroscopic dynamics of the KdV soliton gas predicted by the kinetic equation for solitons with the results of direct numerical simulations of the KdV equation. The simulations involved $200$ solitons enabling an accurate determination of macroscopic parameters of the soliton gas. Two test problems have been considered: the propagation of a trial soliton through a one-component `\emph{cold}' soliton gas and the shock tube problem involving the interaction of two cold gases with different parameters leading to the formation of an incoherent dispersive shock wave. In both cases the excellent agreement between the asymptotic analytical predictions of the kinetic equation and the direct KdV `\emph{molecular dynamics}' numerical simulations has been observed. This confirms validity of the kinetic equation for solitons as a quantitatively accurate model for the description of non-equilibrium dynamics of soliton gases in integrable systems. The challenging problem is now to study the structure and evolution of the definitive statistical characteristics  of integrable soliton turbulence (PDF, power spectrum density \etc). This will be the subject of future work.


\subsection*{Acknowledgments}
\addcontentsline{toc}{subsection}{Acknowledgments}

G.~\textsc{El} is grateful to A.~\textsc{Kamchatnov} for numerous stimulating discussions.
\smallskip


\addcontentsline{toc}{section}{References}
\bibliographystyle{abbrv}
\bibliography{biblio}

\end{document}